\documentclass[12pt]{article}

\usepackage{graphicx}
\input epsf  
 \usepackage{amssymb,wrapfig}  
\usepackage{xypic,cite}
\usepackage[matrix,arrow,curve]{xy}

\def\nn{{\cal N}} 
\def\rr {{\Bbb R}}
\def\cc {{\Bbb C}} 
\def\pp {{\Bbb P}} 
\def\zz {{\Bbb Z}} 
\def\del {\partial} 
\def\cy {Calabi--Yau} 
\def\de {\partial}
\newcommand{\inv}[1]{{#1}^{-1}} 

\makeatletter
\@addtoreset{equation}{section}
\makeatother

\begin{document}

\begin{center}  
                                 
{\hfill SU-ITP-05/09}
  
{\Large \bf{Topological mirror symmetry with fluxes}}  

Alessandro Tomasiello

{\it ITP, Stanford University, Stanford CA 94305-4060}

\end{center}

{\small
Motivated by SU(3) structure compactifications, we show explicitly how to 
construct half--flat topological  mirrors to \cy\
manifolds with NS fluxes. Units of flux are exchanged with torsion 
factors in the cohomology of the mirror; this is the topological 
complement of previous differential--geometric mirror rules. 
The construction modifies explicit SYZ fibrations for compact \cy s. 
The results are of independent interest for SU(3) compactifications. 
For example one can exhibit explicitly which massive forms should be
used for Kaluza--Klein reduction, proving previous conjectures.  
Formality shows that these forms carry 
no topological information; this is also confirmed by infrared limits and
old classification theorems.
}

\section{Introduction}

String theory compactifications on Calabi--Yau manifolds have great
mathematical sophistication but little phenomenological use. Addition of 
fluxes leads to much more realistic alternatives. In these, in general the
supersymmetry is spontaneously broken; dually, one can instead allow 
the manifolds to have SU(3) {\it structure} rather than holonomy. 
Combining the two effects one can get a large variety of 
supersymmetry breakings, as well as some cases in which some supersymmetry
is preserved by the vacuum.
Unfortunately, in many respects common SU(3) structure manifolds 
have so far revealed much less complexity 
than their more prestigious Calabi--Yau subclass. Maybe string theory can 
change that.

Questions about low--energy theories on a \cy\ are answered clearly in 
terms of its topology and algebraic geometry, without ever writing (or being 
able to write) metrics. On SU(3) structure manifolds, can we still write the 
spectrum and some of the interactions? what depends on the topology and what
on our choice of a metric or of a different structure? 

Let us start from what is known.
First of all, for compactifications of type II theories
it has been argued in \cite{glmw} that existence of SU(3) structure 
is enough to ensure that  the
four--dimensional action have $\nn=2$ supersymmetry.
Basically this is because we want a 
well--defined internal spinor to reduce supersymmetry transformations from ten 
to four. Remember that SU(3) structure is just a topological condition; it
can be rephrased as ${\cal W}_3=0$ (a Stiefel--Whitney class) and $c_1=0$.  
Of course such a mild requirement will not guarantee anything on vacua: 
as we said, in these theories supersymmetry will be spontaneously broken. 
(Something 
similar happens when we introduce fluxes but we force the manifold to stay
\cy.) But we can already hope to use the power of supersymmetry in some way. 

Let us turn for a moment to a related problem: that of classifying  
$\nn=1$ vacua. One can
draw a correspondence between fluxes of type II and the 
differential--geometric 
classification of SU(3) structures, which gives some kind of handle on the 
geometry, a technique demonstrated in many papers (starting from 
\cite{gmpw}). 
In the ensuing jungle of possibilities, somehow elegantly \cite{gmpt} 
the manifolds are always either complex or symplectic, making one hope that 
one could use topological strings. Note also that vanishing $c_1$ was already 
assumed above. Moreover, the NS fields come in a form suggestive of mirror
symmetry. 

In fact, there may be a deeper reason for this mirror symmetry to appear. 
Usual \cy\ mirror symmetry is inferred by comparing supersymmetric sigma 
models which correspond to vacua of the four dimensional effective theories. 
If we have spontaneous symmetry breaking, we do not know anything about vacua. 
But in any theory there are many solutions other than vacua. It is enough
to give up independence on spacetime; we can in this way realize domain walls
and other defects, which can a priori be supersymmetric. 
Much in this spirit, \cite{glmw} found evidence for mirror symmetry in 
certain theories with spontaneous supersymmetry breaking. Some of these 
theories were obtained from compactifications on manifolds of SU(3) structure
but not \cy. Later \cite{fmt,gmpt} elaborated further on this mirror symmetry.

These are encouraging results. They have been obtained, however, 
by somewhat indirect arguments, and in particular without knowing 
 what is the general four--dimensional theory obtained by compactification on 
a manifold of SU(3) structure. For example, even the example considered above
involved certain conjectures (if reasonable ones) on the spectrum of the 
Laplacian 
on the internal manifold (or, as we will argue, on the signature operator).

I will take here some first steps in that direction. I will consider 
a concrete example and then try to take advantage from it to infer some
general ideas. These are roughly speaking the contents respectively of section
\ref{sec:mirror} and \ref{sec:gen}. 

What is called here an ``example'' may disconcert some. I do not give any
metric or detailed expressions on the SU(3) structure. I concentrate instead
on the topology, and on some indirect differential--geometric properties. 
I explain along the way why this is enough to make some steps forward; a 
 paradigm to keep in mind is of course the \cy\ case, in which no metric
is ever required. In the present context this is less obvious, as we may seem
to require information on the spectrum of the Laplacian, for example. 

The class of examples is basically making more concrete the ones 
considered in \cite{glmw}. The strategy is to modify certain 
SYZ constructions previously introduced by several mathematicians. 
These constructions are reviewed in detail, and straightforwardly elaborated 
on, in section \ref{sec:gross}; much of the attention will be restricted 
to the case of the quintic, to fix ideas. Moving on, in section 
\ref{sec:mirror} we explain why and how the mirror quintic gets modified.
 We will 
be able to identify in certain torsion groups the mirror of the integral part
of the NS fields (the flux). It will be 
interesting to compare this with previous results \cite{fmt} 
about the differential--geometric (as opposed to topological) part 
of mirror symmetry. 
In the final section I will try to infer general statements about 
the structure of the four--dimensional theory, in particular trying to 
justify a posteriori some of my statements above, 
that explicit knowledge of the metric should not be required in
order to deal with compactifications on manifolds of SU(3) structure. 
One of the points in this
argument is the topological classification of six--manifolds, which,
as we will remark, 
fits nicely with the data needed to define an $\nn=2$ 
supergravity.

\section{Topological mirrors}
\label{sec:gross}

In this section we will review
a topological construction of SYZ fibrations for \cy\ manifolds
due to various authors \cite{gross,ruan,zharkov}. 
We will mainly follow ideas in \cite{gross},
emphasizing however aspects that we will need later. 

\subsection{Review of the SYZ fibration for the quintic}

The SYZ conjecture \cite{syz} says that every \cy\ is a special Lagrangian 
$T^3$--fibration with
singular fibres, and that mirror symmetry is T--duality along the fibres. 
It is a consequence of including branes in mirror symmetry.
It is still unproven; but if we forget about the
special Lagrangian requirement 
it is by now very well understood. The quintic is the example which has been
considered first, the clearest one and hence the one we will consider here. 
It is understood that most of what we say has generalizations to complete 
intersections in toric manifolds. 

Rather than merely recite the construction, we will try to justify it in 
steps, skipping or sketching details along the way. 

First of all, 
it is easy to see that the base $B$ of the fibration has always $H^1(B,\rr)=0$;
the argument is similar to the one that establishes that the fibre is $T^3$. 
The mirror of 
a brane on the base should be a brane on the whole mirror \cy; this brane is 
rigid (it is ${\cal O}_{CY}$), and hence also the original should be rigid. 
Deformations of a special Lagrangian brane are given by its $H^1$, which gives
the desired result.

For the quintic there is a natural way to see what this base is (which is
in fact a $S^3$). Remember
that it is the locus of a quintic polynomial in $\cc\pp^3$. In toric geometry, 
the polytope representing $\cc\pp^3$ is a four--dimensional ``penta--hedron'' 
whose five faces $F_i$, $i=1\ldots 5$ are each a tetrahedron. One way of
seeing this pentahedron is as base space of a $T^4$ fibration; on each face 
the fibre degenerates to $T^3$ in various ways.  
It is standard that the face $F_i$ can be thought of as the locus 
$\{ z_i=0\}$, where $z_i$ are the five homogeneous coordinates. 
Suppose we take as a particular quintic polynomial $z_1 z_2 z_3 z_4 z_5=0$. 
This consists of the boundary of the pentahedron, which is topologically 
the boundary of a four--ball, that is an $S^3$. Moreover, as we said, on 
this boundary the fibre of the toric fibration is at most three--dimensional.

We are not done for several reasons. First of all, the quintic we have 
considered is a singular one: all the intersections $F_i \cap F_j$ 
are singularities. Second, on these loci the fibre of the toric fibration 
drops of dimension again, becoming a $T^2$. 

Both problems can be solved by deforming slightly the polynomial we had taken,
say by adding a small Fermat piece, $\epsilon (z_1^5 + \ldots + z_5^5)$. 

\begin{figure}[h]
  \centering
\includegraphics[angle=0,width=4in]{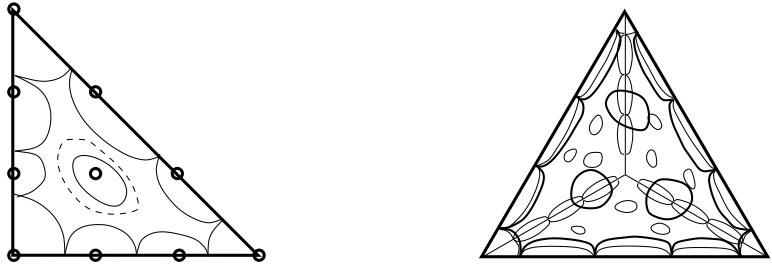}
  \caption{\small Amoebas in one and two complex dimensions.}
  \label{fig:amoebas}
\end{figure}

To understand what will happen, we can take inspiration from lower--dimensional
examples, for example a cubic in $\cc\pp^2$ (which is called 
an elliptic plane curve). A well--known way to visualize this Riemann surface 
uses so--called {\it amoebas}, introduced in \cite{gkz}. For an equation 
$f(x,y)=0$ in $\cc^2$, this would be the image of the locus under the image 
$(x,y)\mapsto (|\log(x)|, |\log(y)|)$. In our case, essentially 
we can  perform 
this map in each of the usual three patches of $\cc\pp^2$ and glue the 
results. The result for a cubic curve is the interior of the solid lines in 
the triangle in figure \ref{fig:amoebas}. It is not hard
to imagine a Riemann surface of genus one fibred over the amoeba inside the
triangle. 

The fact that there is one hole corresponds to the fact that there is one 
internal point in the toric diagram (the equation is cubic) and to the fact
that the resulting Riemann surface is of genus one. If we had taken a 
polynomial of degree $d$, there would have been ${d-1 \choose 2}$ points
in the interior of the toric polytope; this number would have also been 
the genus of the curve. 
 These facts all generalize from $\cc\pp^2$ to other toric Fano manifolds 
(Hirzebruch etc.): the genus of a generic curve inside the Fano  
is always equal to the number of internal points, choosing the size of 
the polyhedron according to the degree of the embedding.
There are many amusing proofs of these facts but we will not 
concentrate on this. 
Rather, let us go back to our fibrations: we can imagine 
a spine inside the amoeba, shown as the dashed line in figure 
\ref{fig:amoebas}. 
The Riemann surface can be $S^1$--fibred over this spine; the spine can be
imagined as a deformation of the boundary. Had we taken, as for the quintic, 
a polynomial $z_1 z_2 z_3=0$ for the elliptic curve, the amoeba and its spine
would have been pushed to the boundary of the triangle. 

Of course it becomes harder to imagine all this in higher dimensions. Already
imagining a quartic in $\cc\pp^3$, which is of course a K3, becomes less easy;
on the right of figure \ref{fig:amoebas} 
we have attempted a messy drawing of the intersection of the amoeba with 
the boundary of the toric polytope. Note that the intersection with each 
face is now a quartic in $\cc\pp^2$, which has genus 3, so we have 
three ovals per face. We can  convince
ourselves that things cannot be as easy as before by considering that now
the ``two--dimensional spine'' would be an $S^2$; a $T^2$--fibration over
an $S^2$ cannot give a K3. We know there have to be singular fibres. 
Where are these singularities occurring? a natural candidate answer is: where 
the amoeba meets the actual boundary of the toric polytope. One feature that 
this boundary did not have in the previous case was faces of codimension two:
edges. The amoeba will meet each of these four times because the
polynomial is of dimension four, just as in the previous case it was meeting
each side three times. There are six edges in a tetrahedron. This gives a 
total of $4\times 6=24$ such points. And it is well--known that this is 
the number of singularities of a $T^2$--fibred K3. 

\begin{figure}[h]
  \centering
\includegraphics[angle=0,width=6in]{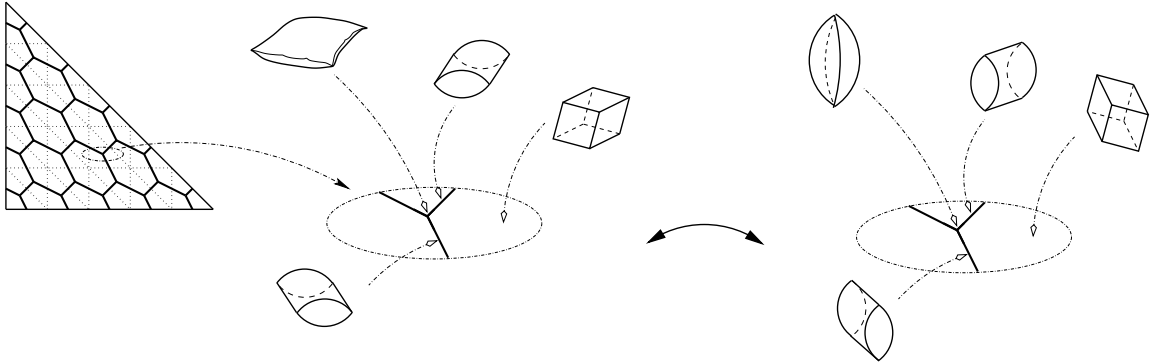}
  \caption{The discrimimant locus $\Delta$ and the singular fibres over it. 
Also shown are the regular $T^3$ fibres over points outside $\Delta$.}
  \label{fig:deg}
\end{figure}

This gradual growth in dimensionality should hopefully make it now
easier to imagine what will happen in complex dimension 3. The toric polytope 
is now a four--dimensional pentahedron; degenerations happen again on the 
intersection with the faces of codimension two, which this time are triangles. 
This intersection is now a quintic in $\cc\pp^2$, which has genus 6. 
The singular fibres lie on the solid graph in the triangle 
in figure \ref{fig:deg}. (There are ten such two--faces in a pentahedron.) 

The picture shows the fibres over various points. Outside the graph, call
it $\Delta$, the fibre is a three--torus, which we show as a cube on which
we have to imagine opposite sides identified. $\Delta$ is also
called discriminant locus. On edges of $\Delta$,
a $T^2$ degenerates to an $S^1$, but the whole fibre stays three--dimensional. 
(This is to be contrasted with toric fibrations, in which fibres drop 
of dimension on discriminant loci.)
It is not hard to see that these singular fibres $F_{\mathrm edge}$ 
are an $S^1$
times an $S^2$ with two points identified. The latter two--dimensional 
factor is the singular fibre which we had in complex dimension two, 
for the K3. In the picture it is also shown how the $T^2$ which degenerates
is related to the direction of the line in the graph $\Delta$ over 
which it is a fibre. 
Over a {\it vertex} of $\Delta$, the 
fibre $F_{\mathrm vertex}$ is where both $T^2$'s degenerate, giving rise
to a cushion--like shape, again with its upper and lower faces identified.

Actually, these are not the only types of singular fibres. Where 
$\Delta$ meets the edges of this triangle, three two--faces actually meet, 
and we have another type of vertex. The fibre over those vertices degenerates
differently. We are not going to make much use of those vertices, and 
anyway the situation around those is identical to what happens after
T--duality around the vertex already considered, so we move on to 
consider the T--dual vertex. 

This is what is depicted on the right of figure \ref{fig:deg}. 
Over a point outside
$\Delta$, the fibre is just the dual three--torus. What happens to the
other fibres? We have to figure out what happens to T--duality as
some cycles in the torus shrinks. On first thought one may think that
the dual cycle will grow noncompact. Rescaling to keep those finite 
does not help, because all the rest would be then rescaled to zero. 

The answer comes from considering monodromies around $\Delta$. 
Not surprisingly, these monodromies are in correspondence with the
topology of the singular fibres. One knows how monodromies transform 
under T--duality, and hence one can infer the topology of the dual fibres.
Essentially what we are doing is, doing T--duality as we know it outside
the singular locus, and then putting back in the singular fibres in such 
a way as the resulting space to be a manifold. 

The singular fibres $\tilde F_{\mathrm edge}$ 
that one finds in the T--dual over an edge are of the same type as 
the original ones, $F_{\mathrm edge}$, except the $T^2$ which 
degenerates is a
different one. Looking at the common degenerating cycle of the two 
$\tilde F_{\mathrm edge}$ this time gives a different $F_{\mathrm vertex}$:
it has the shape that one would get from the intersection of two solid
cylinders, and again we have to keep opposite sides identified. 

This pictorial description of the discriminant locus will be useful (and maybe
clearer) in describing what the non--trivial cycles of the fibration are, which
we will do in next subsection. 

Before doing that, let us mention that, other than the SYZ procedure 
we have just described, there is also a more traditional way of
thinking about the mirror in this context: the approach by Batyrev 
\cite{batyrev}. The first remark we can do
is that a hypersurface in a toric manifold
is \cy\ iff the number of lattice points internal to it is one. 
(Notice that this dovetails nicely with the criterion given above for 
the genus of 
curves: in complex dimension one, a \cy\ is simply a curve of genus one.)
Now, we can project rays from this internal point to the polytope and get
a collection of cones of various dimensions. This is not the toric fan of 
the manifold but what is called its dual fan. What is the toric fan itself?
it turns out that it is the dual toric fan of the mirror manifold. In other
words, 
Fan$_{\widetilde M}=\widetilde {\mathrm {Fan}}_{M}$, where the tilde means 
``dual'' (in the toric sense) on Fan, and ``mirror'' on $M$.

\subsection{Cycles}

If we believe this is the quintic, we must see somewhere in this fibration
the famous 204 three--cycles, the one two--cycle and the one four--cycle. 
And once T--duality has been performed we must see that these cycles have
been exchanged. What follows is a straightforward distillation of 
the mathematical literature, in particular \cite{gross,gross1}. 

If the fibration were trivial and without singular fibres, one would
have three one--cycles coming from the fibre, three two--cycles again coming 
from the fibre, 
one three-cycle from the base and one from the whole fibre. This is quite
far from any \cy\ we may be interested to study (for example it is far
from being simply connected). Keeping fibres non--singular we cannot make
the fibration non--trivial: the base has no cycles on which we may 
want to introduce monodromies, or no two--cycles on which we may want
to have a non--trivial Chern class. (We will come back on the distinction
between the two later.) Having introduced singular fibres the situation
changes drastically: the one--cycles become trivial because they all 
shrink on some portion of $\Delta$, as we have seen. Where are the
two-- and three--cycles?

We will first give a pictorial answer and then translate it into a 
spectral sequence that will be useful later. The basic
idea has already surfaced in many context, and often also in physics:
the projection of a cycle to the base of a fibration with singular fibres
need not be a cycle (which would be bad in our case, since there are none)
but can also be a chain (not necessarily closed under the homology boundary 
operator on the base). For example we can take a path (not a loop) 
in the base, joining two cushion singularities. If we choose a random 
cycle in the fibre, the resulting chain will have a boundary: the 
fibre over the endpoints of the path. But if the cycle in the fibre shrinks to
something of smaller dimensionality on the endpoints, there is no boundary, 
and we have a cycle. 

\begin{figure}[h]
  \centering
\includegraphics[angle=0,width=6in]{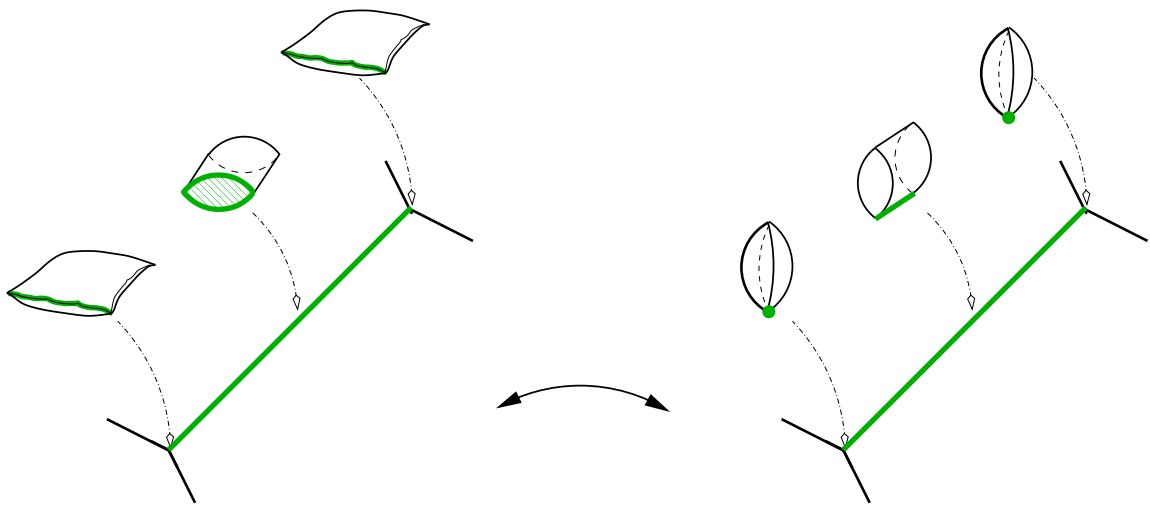}
  \caption{\small 
In green, a three--cycle which projects to a path (left), and 
its mirror (right).}
  \label{fig:path}
\end{figure}

For example, on a path from two cushion singularities, we can have a 
three--cycle. The picture shows how the cycle is wrapped on the fibres: as
we said, the trick is that the dimension of this cycle drops. We have chosen
a path which lies in $\Delta$, but that is not essential. Dualizing this type 
of cycle produces a two--cycle in the mirror as shown on the right of
the picture. 

Another type of possible three--cycle is one in which the open chain in the 
base manifold is a ``plaquette'' of dimension two. It may for example fill
one of the hexagons of the diagram in figure \ref{fig:deg} 
and end on it. In figure 4
we show some of the degenerations involved; once again the idea is that
the cycle in the fibre shrinks at the boundary. As also shown in the figure, 
the dual of this cycle is a four--cycle in the mirror quintic. 

\begin{figure}[h]
  \centering
\includegraphics[angle=0,width=6in]{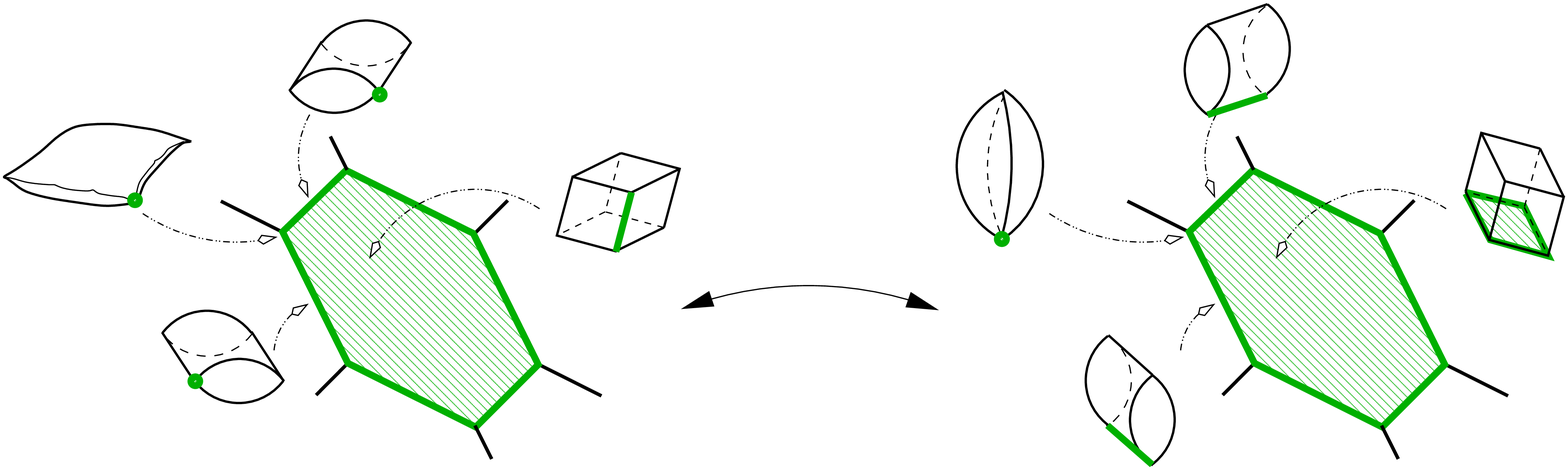}
  \caption{\small In green, a 
three--cycle which projects to an hexagon (left), and its mirror (right). }
  \label{fig:plaq}
\end{figure}

This way of producing cycles is standard. 
It is also instructive to look at the mirror cycles using the Batyrev
construction we briefly described above. 
The mirror quintic is a hypersurface of a toric manifold whose 
polytope is dual to the polytope of the polytope of $\cc\pp^4$. 
It is again
a pentahedron, but bigger. (To gain intuition about this one can again 
consider a lower dimensional example, for example the dual to the $T^2$ 
on the left of figure \ref{fig:amoebas}.)
 Unfortunately there will be singularities in the mirror quintic, 
and to dispose of them we have actually to blow up the mirror polytope 
several times. Rather than trying to visualize this, we can remember that
blowups correspond to adding one--dimensional cones (rays) in the toric fan.
The toric fan and the dual toric fan are exchanged by mirror symmetry, and
this means that the possible blowups of the mirror quintic can be seen 
as adding points in the boundary of the toric polytope for the original 
quintic; the one--dimensional cones will be joining the barycenter of the 
pentahedron with these points. 
These points are nothing else than the centers of the hexagons in figure 
\ref{fig:amoebas},
including those on the boundaries of the triangle. Let us count them. 
The internal ones are six per face. The ones on the boundaries are four per
edge. These give $6\times 10 + 4\times 10=100$. Then there are the five 
original vertices. This gives a total of 105 vertices in the dual 
toric polytope
of the mirror quintic. It is known that one--dimensional cones in the fan, 
or vertices in the dual toric polytope, are in correspondence to 
divisors (holomorphic cycles of complex codimension one) of the manifold;
moreover, there are as many relations among them as the dimension of the
polytope. In this case this is four, so we have our $105 - 4 = 101$ divisors.

Among these, focus on the divisors coming from the centers of the hexagons 
completely internal to one face. These four--cycles are nothing but
the ones depicted on the right of figure \ref{fig:plaq}, filling 
up the hexagon in the base. 
Also, consider the intersection of two neighbouring divisors. This is 
a two--cycle and is nothing but the cycle on a path depicted on the right
of figure \ref{fig:path}. 

\subsection{Leray--Hirsch spectral sequence}

It will turn out to be useful to rephrase all this counting in yet another way
(one that will involve some equations, at last). This spectral sequence 
(almost) computes the cohomology of the total space $M$ of a fibration.

As 
usual for spectral sequences, it 
is made up of ``pages'' $E^k_{p,q}$ each of which is an array (hence the
indices $p,q$); we ``turn page'' computing the cohomology of certain 
differentials. From a certain point on the pages all look the same; this
defines a last page $E^\infty_{p,q}$ 
which is related to the cohomology of $M$.
Over $\rr$, it is sufficient to sum the diagonals to obtain the cohomology
of the total space: $H^i(M)=\oplus_{p+q=i} E^\infty_{p.q}$. (Over the integers
the answer is more subtle.)

To be concrete, let us 
start from an easier situation, one that we will consider again later to gain 
intuition about T--duality.
Namely, we could consider a $T^3$ fibration {\it without} singular fibres.
In this case, the second page has as $(p,q)$--th term $H^p(B, H^q(F))$, so it
looks like 
\begin{equation}
  \label{eq:easyss} E_2\ : \quad
  \begin{array}{ccc}
H^0(B, H^3(F)) &\ldots & H^3(B, H^3(F)) \\
\vdots & \ddots & \vdots\\
H^0(B, H^0(F)) &\ldots & H^3(B, H^0(F)) 
  \end{array}\ .
\end{equation}
(I have been ambiguous about whether the coefficients are real or integer, 
on which I will say more later.) Clearly the $(p,q)$--th term in this page
represents forms with $p$ legs along the base and $q$ along the fibre. 
We will continue considering how 
 the situation would evolve from now on for a
general fibration. Then we will specialize to our case $B=S^3$ and 
$F=T^3$, which will be a bit anticlimatic. 

The differential $d_2$ whose cohomology we have to take to pass from 
$E_2$ in (\ref{eq:easyss}) to $E_3$ is essentially multiplication by Euler 
class. To explain this, let us consider for a moment the case $F=S^1$, 
with coordinate $\phi$.  
At the level of differential forms, a vertical form which is well--defined
in the total space is $e \equiv 
d\phi + A$, where $A$ is a connection on the $S^1$ 
bundle over $B$ whose total space we are considering. If we take its
exterior differential, this gives $de= c_1$, where $c_1$ is
the Chern class of the bundle. 
This simple computation can be exploited actually to write down in simple
cases directly the de Rham complex for the total space and compute its 
cohomology (more on this later), and in some sense it can even be lifted
to a computation over $\zz$ (as Chern classes are integral). But in 
the context of spectral sequences, it will give rise to 
a differential $d_2$
\[
\xymatrix{ 
H^i(B, H^j(F)) \ar[drr]^{\cdot c_1} \\
H^i(B, H^{j-1}(F))& H^{i+1}(B, H^{j-1}(F)) & H^{i+2}(B, H^{j-1}(F))}\ .
\]
In the case where $F=S^k$ the analogue of this map involves the Euler class
of the fibration. We can see that an $S^k$ fibration can be non--trivial
over $k+1$--cycles in the base. In the case of a general fibre $F$, all
these maps $d_k$ will be present. One has then to keep computing their
cohomologies.

All this knowledge is certainly not needed in the case $B=S^3$, $F=T^3$. 
Indeed, second page would look like
\[
\begin{array}{cccc}
1 & 0 &0 &1\\ 
3 & 0 &0 &3\\
3 & 0 &0 &3\\
1 & 0 &0 &1
\end{array} \ ;
\] 
a $d_3$ arrow could a priori connect the first and the last column, but
there is no second Euler class in this case. 

Clearly the resulting cohomology has little to do with the one of a typical
\cy, for example neither with the quintic nor with its mirror that we 
are considering in this paper. So obviously the simplifying 
assumption we have made above, that the fibres are all isomorphic, is 
unrealistic. Fortunately there is a version of the Leray--Hirsch
spectral sequence that can be applied to cases with singular fibres.  
Rather than the groups we were considering above, this one involves 
sheaves obtained from the right derived functor of the projection: 
the resulting groups are denoted $H^p(B,R^q \pi_* \zz )$. 
The second page in our case looks like 
\begin{equation}
  \label{eq:ss}
E_2\ : 
\begin{array}{cccc}
\zz & 0 &0 & \zz \\ 
0 & H^1(B, R^2 \pi_* \zz) &  H^2(B, R^2 \pi_* \zz) &  0 \\
0 & H^1(B, R^1 \pi_* \zz) &  H^2(B, R^1 \pi_* \zz) &  0 \\
\zz & 0 &0 & \zz
\end{array}  
\end{equation}
Without going through an explanation of the definition of the sheaves involved,
we can simply notice that the four non--trivial terms in this sequence 
are the Poincar\'e duals of the cycles we have depicted in figures 3 and 4, 
with the understanding 
that $H^p(B, R^q \pi_* \zz)$ is dual to a cycle with $3-p$ legs in the base
and $3-q$ legs in the fibre. 

We could now wonder whether there will be any map $d_2$ in this $E_2$ page. 
Roughly speaking, the answer is no because for \cy\ manifolds the $T^3$ 
fibrations have monodromies, but no Chern classes. So the cocycles in 
(\ref{eq:ss}) count actually the cohomology of the \cy. 
The situation becomes
different when we consider manifolds different from \cy 's. We now proceed
to explain both facts in more detail.

\section{Mirrors to NS flux}
\label{sec:mirror}

We will start here from some remarks on $T^3$ fibrations, and on the role 
in them of Chern classes and monodromies. Then we will introduce $H$ flux, 
and proceed to T--duality to find the topology of the mirror. This will be
a certain modification of the mirror quintic.

\subsection{Monodromies and Chern classes}

Consider a $T^3$ fibration on a manifold. We will exclude for the time being 
the singular locus $\Delta$, hence considering the fibration restricted to 
$B_0\equiv B- \Delta$. How do we classify such a fibration? first of all 
we have to identify what is the structure group $G$, the group in which 
transition functions take value. One might think that, since
the fibre is already a group, we can consider it as a principal fibration 
and consider $G=U(1)^3$. These would be classified by three Chern classes
in $H^2(B_0, \zz)$. It is clear, though, that this is too restrictive. 
In this way we would exclude the possibility that for example two of the 
$S^1$ be exchanged by a monodromy: this is not possible for a principal 
fibration. The problem is that with this choice $T^3$ acts on itself only
by translations; what we need is to consider $G$ to be the group of 
automorphisms
of $T^3$, which includes rotations and translations. Rotations are relevant
over closed paths in the base; translations (being associated to Chern 
classes) are relevant over two--cycles. We should actually be more careful
about calling them Chern classes: once  
we assign monodromies, the fibration is no longer principal and it is no
longer obvious that we can associate to it Chern classes. The answer
is known: in a situation with monodromies we can define 
a twisted cohomology group $\tilde H^2(B, \zz)$ and Chern classes take 
values in it. 

To make this abstract discussion clearer, 
let us look at the metric on a $T^3$ fibration. This has
the form 
\begin{equation}
  \label{eq:metric}
  g= g_{ij} dy^i dy^j + h_{\alpha\beta} e^\alpha e^\beta\ ;
\qquad e^\alpha\equiv d\phi^\alpha + A^\alpha \ ,
\end{equation}
where $y_i$ are coordinates on the base, $\phi^\alpha$ are on the fibre 
(so both indices $i$ and $\alpha$ run  from 1 to 3) and 
$A^\alpha$ is a connection form, . Its curvatures $c_\alpha$ are the first 
(twisted) Chern classes of the fibration.  Monodromies affect the 
metric on the fibre, $h_{\alpha\beta}$ as $h\to M^t h M$. 

To recapitulate, for a monodromy, pick a one--cycle $S^1$ in the manifold;
after a full loop the fibre undergoes
\[
\left(\begin{array}{c}
\phi_1 \\ \phi_2\\ \phi_3
\end{array}\right)
\ \to \  M 
\left(\begin{array}{c}
\phi^1 \\ \phi^2\\ \phi^3
\end{array}\right)\ .
\]
For Chern classes, on the contrary, pick a {\it two}--cycle, say a sphere;
divide this sphere in two disks $D^2$ and $\widetilde D^2$; then the fibre on 
the two discs is attached to the other via a translation:
\[
\left(\begin{array}{c}
\widetilde \phi_1 \\ \widetilde \phi_2\\ \widetilde \phi_3
\end{array}\right)\sim 
\left(\begin{array}{c}
\phi^1 \\ \phi^2\\ \phi^3
\end{array}\right) + \theta
\left(\begin{array}{c}
c^1_1 \\ c^2_1\\ c^3_1
\end{array}\right)\ .
\]
Here $\theta$ is a coordinate on the equator of both $D^2$ and $\widetilde
D^2$; $\widetilde \phi^i$ are coordinates  on the fibre over 
$\widetilde D^2$, and $\phi^i$ on the fibre over $D^2$. 
 
Summarizing, monodromies are rotations, Chern classes are translations. 

\subsection{T--duality, topology and SU(3) structures}

Suppose now we introduce $H$ flux. The $B$ potential can in general 
be decomposed in fibre--fibre, fibre--base and base--base pieces.  
If we now perform T--duality, these various components will have very different
behaviour. The fibre--fibre component is the nastiest one: it will give rise
in the mirror to non--geometrical backgrounds, that is, to backgrounds in 
which transition functions are valued in O$(d,d)$. One can easily realize
looking at some example that naive application of Buscher rules will give
metrics which do not seem to be well--defined on the dual torus. 
There are actually also other mathematical interpretations of these T--duals
(see for example \cite{bhm}). 
We will however ignore this situation in the present paper, 
and take a B--field of the form
\begin{equation}
  \label{eq:B}
  B_2 = \frac{1}{2} B_{ij}\, dy^i \wedge dy^j 
         + B_\alpha \wedge (d\phi^\alpha + \frac{1}{2}\, \lambda^\alpha)\ .
\end{equation}
It is now easy to perform T--duality along $T^3$. The computation is 
standard and has been done many times; here are the result, which are
also reviewed in some more detail in \cite{fmt}. 

It turns out that the result can be summarized by the rules
\begin{equation}
  \label{eq:T}
    h \longleftrightarrow \inv h \ ; \qquad
  B_\alpha \longleftrightarrow \lambda^\alpha\ .
\end{equation}
The second part of this equation is particularly interesting. Differentiating 
it we get $c_1^\alpha \longleftrightarrow H^\alpha$, where remember that
$c_1^\alpha$ are the three (first) Chern classes of the fibration, 
and $H^\alpha$ are 
the three components of $H$ with two legs along the base. Changing the 
Chern classes changes in particular the topology of the total space, a fact
which has been known to happen under T--duality for some time, starting
from \cite{ag}, later in \cite{duff}; more recently reexamined in 
\cite{kstt} as an early example of SU(3) compactifications (see also 
\cite{schulz} and 
in \cite{bem} in presence of branes. 

In \cite{fmt} this was reconsidered to determine the differential geometric
properties of SU(3) structure manifolds under T--duality. 
Here we are going to consider how the topology changes. Let us expand on 
the difference between the two. 

An SU(3) structure on $M$ 
is a reduction of the structure group of the tangent bundle $TM$ 
to SU(3). That is, it is the possibility to find transition functions which 
are valued in SU(3) rather than in Gl(6). Concretely, this is possible when 
one of the following 
two objects is defined on $M$: i) a spinor without zeros; ii)
a pair $(J, \Omega)$, where $J$ is a real two--form, $\Omega$ is a complex
three--form, both are non--degenerate and they satisfy $J \Omega=0$, 
$i \Omega \bar \Omega = J^3$. The existence of the spinor or of the pair
is equivalent. As we are going to see later, these SU(3) structures are 
important for supergravity compactifications. For this reason, establishing
mirror symmetry with flux requires asking how SU(3) structures transform
under three T--dualities. This question was considered in \cite{glmw, fmt};
in \cite{fmt} the transformation rule was interpreted in terms of
the exchange of two Cl(6,6) pure spinors $e^{i J}$ and $\Omega$, 
objects which are important in generalized complex geometry. 

The mere existence of an SU(3) structure, however, is a purely topological
condition. One would like to know what are the differential properties
of those forms $J$ and $\Omega$. For example, is $J$ integrable? is it closed?
It turns out that their full differential content is expressed by their 
exterior differentials $dJ$ and $d\Omega$. These tensors are collectively 
called {\it intrinsic torsion}. For example, integrability of
$J$ can be expressed as $d\Omega_{2,2}=0$. Often $dJ$ and $d\Omega$ are
 also decomposed in SU(3) representations, which gives rise to tensors
$W_i$, $i=1\ldots5$ \cite{gray}; we will however not need these explicitly. 

More precisely, one can expand $dJ$ and $d\Omega$ in components;
the work done in \cite{fmt} consisted in computing how these components 
went in one another under T--duality. That such a mapping must exist
is obvious if we take into account the fibration structure. Indeed, in terms
of the SO(3) defined by the base, the intrinsic torsion decomposes 
conveniently into two $1$, two $5$ and two $3$, which are exchanged by 
T--duality.

It is less obvious, however, 
that such a mapping can be reexpressed without any reference to the base or
the fibre, as is the case:
\begin{equation}
  \label{eq:ms}
  (\nabla J + H)_{ijk} \longleftrightarrow (\nabla J - H)_{i \bar j \bar k}\ .
\end{equation}
Here we do not have to make any reference to the 1, 5 or 3, which explicitly
contain the information of the presence of the base. 
(For details, see \cite{fmt} or \cite{gmpt}.) 

In this computation the topology seems to have been forgotten again. 
One can ask, for example, the following question. $H$ is not just a 
differential form; it is also integral, that is, its periods are integer. 
So it contains some discrete information. What is the mirror of this 
discrete information? We have already given an answer above: it is in 
the Chern classes of the fibration. However, this answer is given in terms
of the fibration structure. This answer is at the same level of the answer 
given before for the T--duality rule for the intrinsic torsions for the
SO(3) representations 1, 5 and 3, which make explicit reference to the 
fibration structure. Is there an intrinsic answer, one 
that does not refer to the fibration structure? It may again be that, 
after having done the dirty work of considering explicitly 
the fibration structure, the result can be distilled in a way usable without
ever knowing about it. The dirty work for the intrinsic torsion was done
in \cite{fmt}; for the topology it is going to be done in the next subsections
of this section, using the material we have reviewed above. Without
undue suspense, we can anticipate that the answer is in some torsion
factor (cyclic groups of the form $\zz_N$) in integral cohomology groups.
To my knowledge this use of the word ``torsion'' bears no relation to 
the ``intrinsic torsion'' above. At least not so far:

Another possible line of thought could have been, looking at
(\ref{eq:ms}), that the mirror of the flux of $H$ should somehow lie in
$\nabla J$. This is a fully legitimate idea, but extracting this integer
part is not as easy as one might think. The only obvious idea to get an integer
is to antisymmetrize to get a form and then integrate over a cycle. 
But the antisymmetrization of $\nabla J$ gives $dJ$, whose cycles are zero. 
($J$ is not a connection on a gerbe, as opposed to the $B$ field; it is a 
genuine two--form.) Hence, we will get that torsion groups should have to do 
with intrinsic torsion, seemingly a pun but maybe not too trivial a statement.

\subsection{Mirror to the quintic with $H$}

In the original \cy\ situation, there is neither $c_\alpha$ nor $H^\alpha$, 
so all  we have to care about is the first part of (\ref{eq:T}). This 
gives the transformation properties of the monodromies which we have
already implicitly used in previous section. Indeed, monodromies 
are related to degenerating cycles, and the latter were illustrated in 
detail. 

The situation becomes more interesting if we now add $H$ flux to a \cy. 
As we said, a case which is more interesting and tractable is the one
in which $H$ has one leg in the fibre and two in the base. This should
give in the T--dual a change of topology.

Let us be more specific and pick $H$ to be the Poincar\'e dual of a cycle
of the type depicted on the left of figure \ref{fig:path}. 
Such an $H$ has indeed two
legs in the base and one in the fibre. What is now its mirror? The circle
in the fibre along which $H$ is becomes, in the mirror, non--trivially fibred:
it should pick a Chern class. (We will see briefly that this is an abuse
of language.)

The way to see how this Chern class arises is the following. Along the path
in figure \ref{fig:path}, take a disc transverse to it and still 
lying in the base. 
Due to the fact that the path ends on the singularities, we have to 
shrink the radius of this disc as we approach them, so the final
result is not really (the closure of) a tubular neighbourhood, but
more a ``rugby ball'' neighbourhood. 

Now, on each of  these discs it is clear to what the Chern class should
give rise, with a little extrapolation from known cases. We will first
describe what is the procedure just by analogy with known cases, and then
justify it with increasing precision. 
Let us start with an example: the most familiar case of a 
nontrivial Chern class, the Hopf fibration $S^1 \hookrightarrow S^3/\zz_N 
\to S^2$.
This can be obtained by attaching two copies of $D^2 \times S^1$ with a
transition function; if $\theta$ is an angular variable which runs on 
the equator of $S^2$, which is an $S^1= \del D^2$, we have to glue the $\phi_N$
on the northern hemisphere to $\phi_S$ in the southern hemisphere as
\[
\phi_N= \phi_S + N \theta\ .
\]
This is a translation of the coordinate $\phi$, as promised above. 

\begin{figure}[h]
  \centering
\includegraphics[angle=0,width=6in]{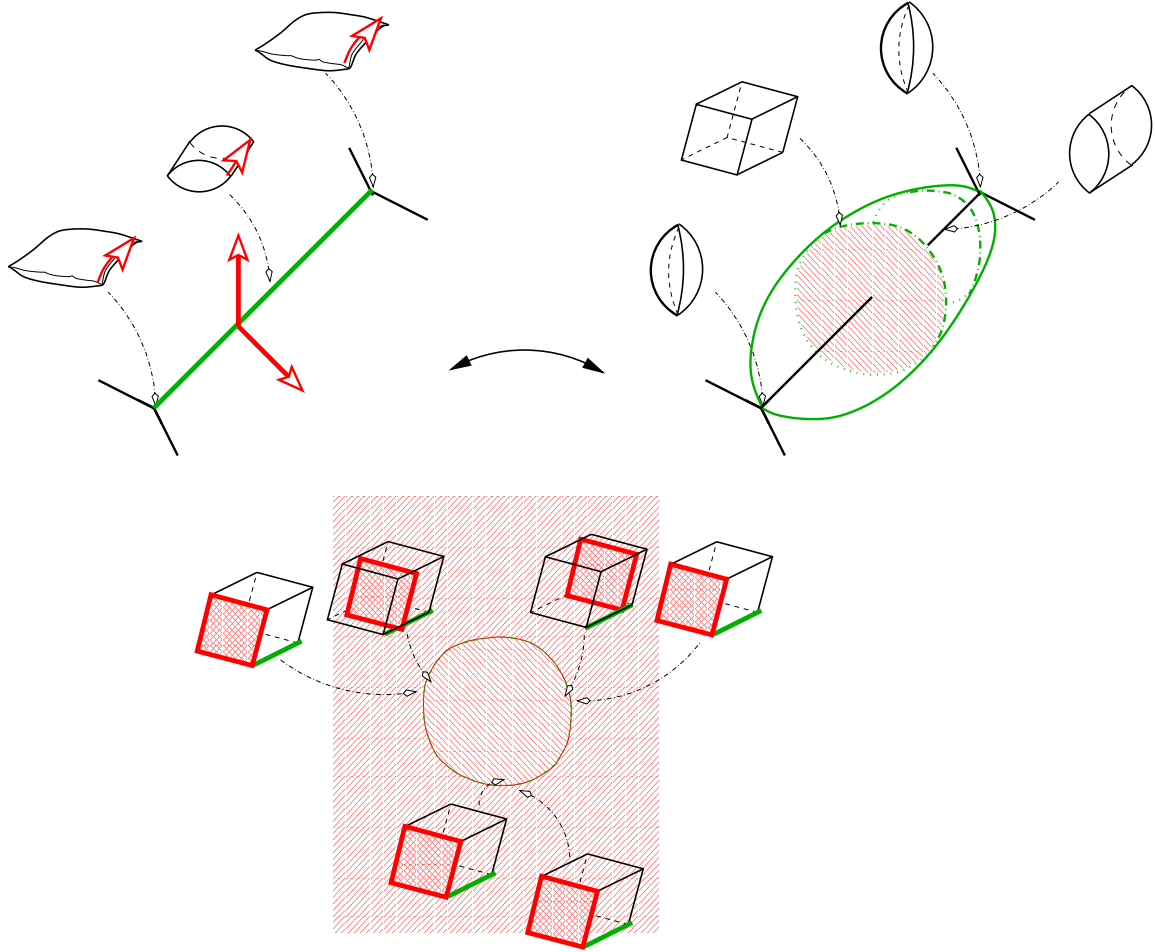}  
 \caption{\small 
The three--form $H$ is switched on along the red arrows in the 
quintic (left). The distorted mirror quintic is defined using a 
``rugby ball'' neighborhood (right). The twisting is on two--chains
in the base which intersects the path transversely. In dashed red is shown
one such a transversal disc.  Along the boundary of this 
neighborhood, the internal $T^3$s are glued
to the external $T^3$s with a twisting of an $S^1\subset T^3$, according
to the rule $\tilde\phi^1 \sim \phi^1 + N\theta$. The locus $\phi_1=$const is 
shown in fibres over the internal and external edges, at three different 
values of $\theta$ (down). The two--chain, in differently dashed red, continues away from
the rugby ball.}
  \label{fig:chern}
\end{figure}

What we have to do now is to perform this 
operation to the $S^1$ shown in the fibre
in figure \ref{fig:chern} 
over each of the transverse discs of the rugby ball neighborhoods. 
So we are detaching the rugby ball with the whole fibre over it, and 
reattaching it after having twisted the $S^1$ as in the Hopf case. 
The number of times $N$ we have to twist is the same as the integer amount
of flux we had in the Poincar\'e dual of the original cycle. This 
``cutting--and--pasting'' operation is similar to (but no the same as) 
what is called usually 
a surgery, and for this reason we will often call it in the following
a surgery by an abuse of language.

We should also come back on another abuse of language: calling this operation
a Chern class. There are no nontrivial classes in $H^2(B)=0$. Why cannot our
twisting be undone? This question is similar to our confusion above, why
the cohomology classes are not those of $S^3\times T^3$. Once again the
cohomologies of the derived sheaves $H^p( B, R^q \pi_* \zz)$ should be 
called to the rescue. The exchange 
$H_\alpha \longleftrightarrow c_1^\alpha$ in (\ref{eq:T}) is still 
literally true, but neither of the two is in $H^2(B)$, again because it 
vanishes. Rather, $H$ lies in $H^2(B, R^1\pi_*\zz)$. It has been shown 
\cite{gross1} that the action of mirror symmetry is to send 
$R^p \pi_* \zz \leftrightarrow R^{3-p} \pi_* \zz$. (This fact is 
clearly consistent with the usual formula $h^{1,1} \leftrightarrow h^{1,1}$.)
So, also $c_1^\alpha$ does not lie in $H^2(B)$, but in 
$H^2(B, R^2 \pi_* \zz)$. This can also be described intuitively. The twisting
we have described is performed over a certain two--chain on the base. 
We have not described where this two--chain ends because it is not important.
Wherever it ends, we have divided it into the disc above and the rest. 
Then we have glued the fibre over the disc to the fibre over the rest with a
twist. We did not need to know what the rest was. But is the twisting really
nontrivial? after all we have said it is a two--chain, not two--cycle (of
which there would be none), in the base. One way of seeing that the 
twisting is nontrivial is that the two discs cannot be moved across the 
singularities at the end of the path. There, the circle we are twisting
becomes cohomologically trivial.  
If we had tried to twist on two-chains which shrank to nothing
on points different from those singularities, we would have run in trouble: 
one would have been forced to shrink the twisted fibre at the end. 
But here the $S^1$ that
we are twisting is cohomologically trivial in the fibre over the singular 
points at the end of the path.

It is maybe not obvious that this gives rise to a manifold again. A sketch
of an argument would be applying prop.~2.7 in \cite{gross}. 
In what follows we will suppose the space we have defined is a manifold and 
call it simply the distorted mirror quintic.

With this explicit description of the mirror manifold we
can now proceed to characterize its topology. We will start by
giving its cohomology, and then we will see that this is actually enough
to characterize completely its diffeomorphism type. 

Before actually considering the distorted mirror quintic, it is maybe 
instructive to look at the cohomology of the Hopf fibration, for which
we know already the answer from a variety of sources. Specifically, we 
want to ask why the would--be Poincar\'e duals of the fibre and of the base
are absent from the cohomology of the total space. As we have 
seen before, a vertical one--form which is well--defined reads $e=d\phi +A$, 
with $A$ a connection such that $dA=c_1=N \mathrm{vol}_{S^2}$.
This one is not in cohomology because $de= N \mathrm{vol}_{S^2}\neq 0$. 
But also, at the level of de Rham cohomology, $\mathrm{vol}_{S^2}$ is in 
cohomology but as a trivial class, since $\mathrm{vol}_{S^2}=\frac 1N de$.
So a Chern class eliminates both a one-- and a two--cycle. If we look 
at integral cohomology, we can actually with some care promote these
forms to (for example) integral cochains of the manifold viewed as 
a CW--complex. Then we can no longer divide by $N$, so $\mathrm{vol}_{S^2}$
is non--trivial but $N\mathrm{vol}_{S^2}= d e$ still is.  This is 
a torsion factor, $\zz_N$. (Summarizing, we have gotten $H^0=H^3=\zz$, 
$H^1=0$, $H^2=\zz_N$.)

Something similar to this toy case will happen for our distorted mirror 
quintic. (There are actually intermediately difficult cases that one could
also analyze, nilmanifolds. The way the cohomology works there has been 
easily worked out in \cite{bem}, along with its K--theory and a proof
that it is consistent with T--duality. So we are not going to review
those cases here.) 
Let us consider for example the two--cycle depicted 
on the right of figure \ref{fig:path} 
for the mirror quintic. Its Poincar\'e dual
is torsion. Indeed, it reads $e^2 e^3 {\mathrm {vol}}_{D^2}$, where
$D^2$ is the transverse disc, and $e^2$ and $e^3$ are the vertical forms in 
the directions unaffected by the twisting. As for the one which {\it is} 
affected by the twisting, we have $d e^1= N {\mathrm {vol}}_{D^2}$. 
So we have $N e^2 e^3 {\mathrm {vol}}_{D^2}= d (e^1 e^2 e^3)$. It is not 
completely obvious, but we can extend again these expressions to 
integral cocycles, as above for the Hopf fibration.

We can see how this modifies the spectral sequence (\ref{eq:ss}). 
The Chern classes we have just described give rise to maps $d_2$ in the 
page $E_2$ we had above:
\begin{equation}
  \label{eq:ssmap}
E_2\ : \qquad
\xymatrix{ 
\zz \ar[drr]^{\cdot c_1} & 0 & 0 & \zz \\
0 & H^1(B, R^2 \pi_* \zz)  &  H^2(B, R^2 \pi_* \zz) &  0 \\
0 & H^1(B, R^1 \pi_* \zz) \ar[drr]^{\cdot c_1}&  H^2(B, R^1 \pi_* \zz) &  0 \\
\zz & 0 &0 & \zz \ . }
\end{equation}
We see here once again that our ``Chern class'' is not actually in $H^2(B)$,
as usually the name implies, but lives in $H^2(B, R^2 \pi_* \zz)$. 
Now, the third page looks like
\begin{equation}
  \label{eq:ss3}
E_3\ : \qquad
\begin{array}{cccc}
0 & 0 &0 & \zz \\ 
0 & H^1(B, R^2 \pi_* \zz) &  H^2(B, R^2 \pi_* \zz)/\zz \oplus \zz_N & 0 \\
0 & H^1(B, R^1 \pi_* \zz)/\zz &  H^2(B, R^1 \pi_* \zz) &  0 \\
\zz & 0 &0 & \zz_N
\end{array}  \ .
\end{equation}
The argument
above for the (Poincar\'e dual of the) two--cycle was direct; at the same
time we have also gotten that one of the three--cycles has gone away (which
we have indicated fancily as a $/\zz$). We now know that the cycles which
are conjugated to these ones should also disappear or become torsion. 
These are a four--cycle and a three--cycle (somehow confusingly, these
are also often called Poincar\'e duals). Seeing what happens to 
the conjugated three--cycle is less direct. We cannot compute directly
the lower $d_2$ map. To do that we should
identify explicitly the map from the disappearing four--cycle. This is not 
easy: the intersection matrix between four-- and two--cycle is somehow
complicated \cite{gross} and without a good idea we are faced with the
task of inverting a 101$\times$101 matrix. However, if we simply put a 
question mark in the lower right corner, we can use Poincar\'e
duality and the universal coefficient theorem (it is a standard technique)
to nail down the remaining group to be $\zz_N$. (I thank Mark Gross for
having pointed this out.)

Were it not for the singularity possibly
created in the procedure above,  
we would have actually described completely the 
diffeomorphism class of the distorted mirror quintic. This is because 
six--manifolds are completely classified by a theorem by Wall and \u{Z}ubr 
which 
we will describe in next section. We will also see that topology is not
the end of the story as far as compactifications of supergravity go; one has
to supplement also some SU(3)--structure information. 

\section{Topology, metric, and compactifications}
\label{sec:gen}

Wall's theorem \cite{wall} classifies completely up to diffeomorphisms 
six--manifolds which are spin, simply connected and whose cohomology
has no torsion.
\u{Z}ubr \cite{zubr} later extended the classification 
to  the case with torsion. The theorem says that there is a 
one--to--one correspondence between such manifolds and the following data:
\begin{itemize}
\item two free abelian groups $H$ and $G$, the latter of even dimension 
(they will be $H^2$ and $H^3$ of the manifold)
\item a symmetric trilinear map $D$ on $H$ (it will be the intersection form)
\item a homomorphism $p_1: H \to \zz$ (it will be the Pontrjagin class)
\end{itemize}
such that $D(x,x,y)= D(x,y,y)$ modulo 2 and $4\mu(x,x,x)=p_1(x)$ modulo
24, for every $x,y \in H$. In the case with torsion, $p_1$ and $D$ get
actually refined and have a more sophisticated definition \cite{zubr}.

The power of this theorem is made possible by the well--known fact that
diffeomorphisms classes essentially do not have any more information 
than homotopy classes from dimension five up. Once we know, thanks to Smale,
that the Poincar\'e conjecture  is valid in dimension bigger
than five (we finally know it is true also in the original three--dimensional 
case \cite{perelman}, 
we can use surgery techniques (similar to the cutting--and--pasting
we performed on the mirror quintic in previous section) to construct manifolds
with any given prescribed invariant starting from a sphere; or, viceversa, 
to start from a given manifold and take away cycles from it until we 
reach a sphere. 

There are various reasons for quoting this theorem here. First of all
to show that we are close to characterize completely 
the topology of our distorted mirror quintic. Second, it has been shown
before that compactifications on this type of manifolds requires consideration
of forms which are not in cohomology, and one could wonder whether these forms
have to do anyway with topology or not; this question is less trivial than it 
looks and we are going to answer it in next subsection. Last, we will 
try to draw from this theorem a more general lesson about the role of 
topology and of differential geometry in string theory compactifications,
which will occupy us in last subsection, there, the theorem by Wall and 
\u{Z}ubr will show up again.

\subsection{A basis for KK reduction}
\label{sec:basis}

We will now see how we are able to recover certain assumptions made in 
\cite{glmw} in order to find the KK reduction of IIA theory on certain SU(3) 
structure manifolds.  
It was argued there that, if one believes that a mirror symmetric 
manifold to a \cy\ with flux had to exist, such a mirror needs to have
a basis of forms which are not harmonic but which satisfy certain 
properties closely related to being harmonic. The reason we care so much 
about that particular case is that we hope to get a handle on 
which mathematical object should replace  
harmonic forms for supergravity compactifications on general (non 
Calabi--Yau) SU(3) structure manifolds. 

The basis comprises two--forms $\omega_i$,  
three--forms $\alpha_A$ and $\beta^A$, and four--forms  $\widetilde \omega^i$.
These forms satisfy: 
\begin{itemize}
\item[{\it i)}] They must be conjugated to each other: 
\begin{equation}
  \label{eq:conj}
  \int \omega_i \tilde \omega^j = \delta_i^{\ j} \ , \quad 
\int \alpha_A \alpha^B = 0\ , \quad \int \alpha_A \beta^B =  
\delta_A^{ \ B} \ , 
\quad \int \beta^A \beta^B = 0\ .
\end{equation}
\item[{\it ii)}] The Hodge $*$ has to close within the basis. 
\item[{\it iii)}] Also the exterior differential has to close. More 
specifically, 
\begin{equation}
  \label{eq:glmw}
  d\omega_i= E_i \alpha_0 \ , \qquad d \alpha_A=0\ , \qquad 
d \beta^A=\delta^{A0}E_i \widetilde\omega^i\ , \qquad d\widetilde \omega^i=0 
\ .
\end{equation}
(This is not the most general way $d$ could close on the basis; one could 
have $d \omega_i$ not all proportional. The most general case has been 
recently advocated in \cite{ferrara}; from there, one reduces to 
(\ref{eq:glmw}) by taking the rank of a certain matrix to be one. We will not
need that generalization here, for reasons to become apparent later.)

As a physical post scriptum, we should add that the $E_i$ can be made 
much smaller than the higher masses of the Laplacian.

\item[{\it iv)}] There exists an SU(3) structure $(J, \Omega)$ whose $J$ is
a linear combination $\sum v_i \omega_i$,
and whose $\Omega$ is a linear combination of the $z^A\alpha_A+ F_A \beta^A$;  
coefficients $v_i$, $z^A$, $F_A$ are {\it constant} on $M$. This 
SU(3) structure satisfies moreover $dJ^2=0$ and $d\mathrm{Re} \Omega=0$ (which
is called {\it half--flatness}). 

This assumption actually {\it almost} follows 
from {\it iii)}. Once nondegenerate $(J, \Omega)$ can be found (which roughly
speaking says that the basis ``covers the whole of the manifold''; we will
come back on this), it is enough to require that $\omega_i \alpha_A=0$ and
$\omega_i \beta^A=0$, which are only vector equations. 
Finally, half--flatness now follows from (\ref{eq:glmw}) after rotating
$F_0$ to purely imaginary. 

The reason for this assumption about SU(3) structures has  
to do with supersymmetry; we will review this later. 
\end{itemize}

Let us first of all show that such a basis exists on our distorted mirror 
quintic. The idea is to use forms that were already defined on the mirror 
quintic, and simply see how their properties get distorted. More precisely, 
we mean that we can express all the harmonic forms on the mirror quintic
in terms of forms on the basis and of $e^\alpha$'s. The expression obtained
in this way still make sense on the distorted mirror quintic; what has been 
changed is what the $e^\alpha$'s are, or in other words their properties
under the exterior differential $d$. 

Going back to (\ref{eq:ssmap}), we can see that all of the harmonic 
four--forms in the original mirror quintic have two indices along the fibre
and two along the base. Their Poincar\'e duals are of the form depicted 
on the left of figure \ref{fig:path}. For an explicit realization, 
we can take four--forms which are localized around paths, or more precisely
with support on a rugby--ball neighborhood as in figure \ref{fig:chern}.
So they look like $e^2 e^3 \rho vol_{D^2}$, where $\rho$ is a function 
with the support inside the rugby ball, 
and $e^\alpha$ are as in (\ref{eq:metric}). 
When we pass to the distorted mirror quintic, all the four--forms but one
are left undisturbed; the remaining one sits on the path which has been 
affected by the surgery. Let us call $\tilde\omega^i$, $i\neq 1$ all the 
unmodified four--forms, and $\tilde\omega^1$ the modified one. Since $d e^1=
N {\mathrm{vol}}_{D^2}$, we have 
 $d(e^1 e^2 e^3)= N e^2 e^3 {\mathrm{vol}}_{D^2}$. This is one of the 
two maps in (\ref{eq:ssmap}).  
Let us now call $e^1 e^2 e^3\equiv \beta_0$. This shows
explicitly the second two equations of (\ref{eq:glmw}); the other $d_2$ map in 
(\ref{eq:ssmap}) will show the other two. This lower $d_2$ connects the volume 
form of the base (lower right corner), which we call $\alpha_0$, with a 
two--form which is the Poincar\'e dual
of a four--cycle on an hexagon; or more probably a linear combination thereof.
This four--cycle is 
difficult to identify explicitly, for the same reason explained after 
eq. (\ref{eq:ss3}). We know it exists from 
(\ref{eq:ssmap}); call this two--form $\omega_1$. Locally around the 
modified path (which is not necessarily its support however) it looks like
$e^1 {\mathrm{vol}}_{\mathrm{path}}$. 

Summarizing so far, we have essentially claimed that (\ref{eq:glmw}) is, in 
the case of 
the distorted mirror quintic, simply true because of (\ref{eq:ssmap}). 

This shows the mathematical part of {\it iii)}. The physical post scriptum
is more subtle. Computing the Laplacian on the forms gives rise to 
a first massive eigenvalue of order $\sim E^i E^i$. (We are not writing
down explicitly period matrix factors which would take care of the 
Hodge $*$, but by assumption {\it ii)} they give only numbers.) We have to
provide an argument for this first eigenvalue to be much smaller than the
more massive ones. Something very similar is already present in the 
literature. In \cite{dishonesty} it is proven that the spectrum of the 
Laplacian can be made to vary arbitrarily little when performing 
certain surgery operations on a manifold (although it cannot remain 
identical). As stressed in section \ref{sec:mirror}, 
the operation we have performed
to pass from the mirror quintic to the distorted mirror quintic is a 
cutting--and--pasting operation which is not what is usually called surgery;
but it is similar. It should be checked that one can modify that theorem
to fit it to our case, but it is plausible; we will assume this here. 
Then, let us consider the mirror quintic. It has a massless spectrum, and
then a first massive eigenvalue of the Laplacian $m_0$. After our operation,
we certainly have made one of the harmonic forms non--harmonic; but the 
generalization we are assuming of the theorem in \cite{dishonesty} tells
us that we can take the spectrum on the mirror quintic to start with the 
remaining harmonic forms, then have an arbitrarily small first massive 
eigenvalue $\delta m$, and continue with the formerly first massive 
eigenvalue, itself very little modified, $m_0 + \delta m'$. As a result, 
 $\delta m \ll m_0 + \delta m'$, which is what we wanted to show.

The other points are more or less taken verbatim from the 
undistorted mirror quintic. For example, for point {\it ii)} we can observe
that all we have changed in the metric is in the expression of the $e^\alpha$,
not $g_{ij}$ or $h^{\alpha\beta}$. (See eq. (\ref{eq:metric}).) Our basis
has the same expression, in terms of the $e^\alpha$'s, in the distorted and 
undistorted mirror quintic. The Hodge
$*$ has therefore the same action. One can argue similarly for point {\it i)}.
Also, one can look at the local explicit expressions given above. 

There are additional subtleties for point {\it iv)}. Again, we can argue that
the on the distorted mirror quintic $J$ and $\Omega$ have the same linear
expansion in terms of $\omega_i$, $\alpha_A$, $\beta^A$ as $J$ and $\Omega$
on the mirror quintic. This is so far just a definition. Now, we can again 
invoke the fact that the expressions of the forms $\omega_i$, $\alpha_A$ and
$\beta^A$ in terms of the $e^\alpha$'s do not change after our 
surgery operation. Whether $J$ and $\Omega$
define an SU(3) structure is just an algebraic properties; so the $e^\alpha$
are not going to be hit by a differential, and the verification 
is formally still the same as on the mirror quintic. 

Now, this argument is unfortunately a bit abstract: we do not really see
why $J$ and $\Omega$ define an SU(3) structure not even before the 
surgery. This is because we have reviewed in this paper only the topological
construction of the SYZ fibration, and we have not said a word about the
part concerning the fibration being special Lagrangian -- which would 
have involved introducing proper $J$ and $\Omega$. This has not been done
yet and will certainly not be attempted in these pages. 

So, even if we have gotten point {\it iv)}, it is unfortunate that we
do not have a more concrete argument for why there is an SU(3) structure
in the span of forms satisfying (\ref{eq:glmw}). If we had such an argument, 
we could try to apply it to other cases, where the fibration structure
is not coming to the rescue. In what remains of this subsection we will 
present an intuitive argument which does not really work but which 
may be a first step.  
(This is also maybe a place to remind that fortunately, half--flatness has 
already been shown, in a spirit similar to one of the present paper, in 
\cite{fmt}.) 

The argument would go as follows. 
In section 2 we have given, among other things,
a basis for four--cycles, see the right portion of figure \ref{fig:plaq}. 
Each of them is fibred over an hexagon in 
the diagram in figure \ref{fig:amoebas}. 
We can choose the Poincar\'e dual of one such 
a four--cycle to be a two--form localized 
around one of the hexagons. (As a differential form, we can just make it 
fall off smoothly away from the hexagon; as an integral cochain, we would
give an appropriate triangulation of the whole space, and then it would be 
non--zero only on cells including the hexagon.)

We have already admitted our ignorance regarding which four--cycle is lost. 
Suppose for simplicity it is fibred over a certain hexagon (and not a linear
combination). Then, 
an arbitrary combination of all the harmonic two--forms will be
zero over this hexagon. The two--form $J$ appearing in the definition of 
an SU(3) structure has to have no zeros; therefore it could not be a 
combination of harmonic two--forms. The two--form which has disappeared
from the cohomology, which we have called $\omega_1$ above, has 
exactly support on the missing hexagon. 

This argument does not work because our choice for the Poincar\'e duals
is by no means the only one. It is a common choice, but there are many
others; very often one actually takes the Poincar\'e dual of a cycle to 
be defined over the whole manifold. Is there any contradiction in the
possibility that including the first massive form is not actually needed? 
If it were possible to write a 
non--degenerate $J$ from a sum of the {\it harmonic} $\omega_i$ only, it would 
follow that the manifold is symplectic; this may seem to be in contradiction 
with the prediction of half--flatness in \cite{glmw,fmt}. But in fact, one may
well have on the distorted mirror quintic different SU(3) structures with 
different 
differential properties -- one being symplectic or even \cy, the other being
half--flat and mirror to the quintic with $H$. In other words, if we are
considering the distorted mirror quintic because of mirror symmetry, then 
it is natural to consider the SU(3) structure induced by the surgery--like
procedure in section 2; this SU(3) structure will include the massive 
two--form. But a priori we could also decide to compactify on the same manifold
using another SU(3) structure with different properties, should we find one,
and get a different $\nn=2$ effective theory. This discussion opens up 
problems to which we will get back at the end of this subsection. 

\subsection{General SU(3) compactifications}
\label{sec:general}

In this subsection we try to argue general lessons from the particular
manifold we have dealt with so far. The main problem is the right 
generalization of de Rham cohomology to general SU(3) compactifications. 
This will also however raise spin--off problems that we will also try to 
address towards the end; in the process we will also argue that the massless
sector of the theory is in one--to--one correspondence with 
the topology, thanks to the Wall--\u{Z}ubr theorem.

We have shown that forms satisfying
 (\ref{eq:glmw}) exist on the distorted mirror quintic. 
We will now try to understand what is more generally
the meaning of such a basis. Namely, what should replace de Rham cohomology
(or harmonic forms) for compactifications on manifolds of SU(3) structure
which are not Calabi--Yau? As I mentioned before, it is not difficult to
guess a proper generalization of (\ref{eq:glmw}); it has been proposed
in \cite{ferrara}, and is being used in \cite{lgw}. It reads
\begin{equation}
  \label{eq:dftv}
 d \omega_i= a_{iA}\alpha^A - b_i^{\ A} \beta_A \ , \qquad 
d \alpha^A=b_i^{\ A} \widetilde \omega^i \ , \qquad
d \beta_A=a_{iA} \widetilde \omega^i\ , \qquad 
d\widetilde \omega^i=0\ .
\end{equation}
However, whereas for
(\ref{eq:glmw}) we have an explanation now, unfortunately (\ref{eq:dftv})
is a different story. What really counts is the rank of the matrices $a_{iA}$
and $b_i^{\ A}$; it is related in an obvious way to the number of forms
which become massive. In the $T^3$--fibred case, if we go back at 
(\ref{eq:ssmap}), we see that we can only take away two forms from 
$H^3$, the upper--left and lower--right corners. Thus, so far 
we cannot retrieve the case (\ref{eq:dftv}) with a general rank. 
 is in no contradiction 
with anything, since anyway not all SU(3) structure manifolds are
$T^3$--fibred. We will get back later at (\ref{eq:dftv}) later.

So let us try for an alternative. Again, de Rham cohomology cannot be the 
right object, since in it 
$\widetilde \omega^1$, $\omega_1$, $\alpha_0$ and $\beta^0$ disappear.
So they are not part of a basis of harmonic forms. 
We will first examine the possibility that they
have no topological information, answering in the negative, 
and then turn to differential geometry.

The first possibility, that they have some kind of topological information,
is less crazy than it would seem. 
True, Wall classification theorem tells us that hence they are not needed 
in characterizing the topology. But in fact, the de Rham complex has 
often more topological 
information than its cohomology alone. One way to see if it has more
is to strip it to the minimum, so to say -- producing a smaller and manageable
complex called {\it minimal model} \cite{sullivan}. 
This can be used to compute other 
topological quantities (which may be not necessary for a classification, 
though), for example the so--called Massey products, which we are not 
going to review here. Minimal models for nilmanifolds include more than 
just harmonic forms. 
For example, consider an $S^1$--fibration over $T^2$ with $c_1=1$. 
The minimal model for $S^3$ (again) is
generated by the vertical $e$ as defined before and ${\mathrm {vol}}_{T^2}$, 
with the differential acting as we said above, $de={\mathrm {vol}}_{T^2}$. 
So $e$ and ${\mathrm {vol}}_{T^2}$ are in the minimal model even if they 
are not in the cohomology. This clearly looks similar to (\ref{eq:glmw}),
and it is tempting to think that minimal models are the hidden meaning
of those equations. 

Unfortunately this is wrong. A theorem by Miller \cite{miller} implies that
{\it all} simply connected six--manifolds are formal, that is their minimal 
model includes just the cohomology. So it looks like 
the forms $\alpha_0$, $\beta^0$, $\omega_1$ and 
$\widetilde\omega^1$ in (\ref{eq:glmw})
have no topological content. (The other forms, those which survive in the 
de Rham cohomology, obviously have one.) Here we have changed base to one
in which only $E_1\neq 0$.

Having looked at the topology unsuccesfully, we can now turn to the 
differential geometry, that is, at information hidden in the metric -- or more 
precisely, as we will see, in the SU(3) structure.

Having to do KK  reduction on a manifold, the first thing one has in mind to
look for is a basis of forms which are eigen--forms of the Laplacian. 
One could even wonder why this is not enough for 
KK reduction: this would be the generalization of harmonic forms. 

However, while reducing the supergravity action one often needs to know 
the action of the exterior differential and of the Hodge $*$ on forms. 
So knowing that they are eigen--forms of the Laplacian is not enough. 

To this one can easily repair, as is probably known to many, 
by considering the signature operator, $s\equiv
d + d^\dagger$. This is the Dirac operator acting on bispinors, which, as is 
well known, are the same as differential forms. Not surprisingly, it 
squares to the Laplacian: $s^2 = \Delta$. The spectrum of this operator
is related to the one of the Laplacian by a standard construction. Indeed, 
suppose one has 
a form $\nu$ which satisfies $\Delta \nu = m^2 \nu$. Then we can 
construct eigenforms of the signature operator as follows. 
If $(s \pm m) \nu \neq 0$, it is an eigen--form of $s$ with eigenvalue
$\mp m$. If on the other hand one of the two is zero, say $(s - m)
\nu=0$, it means that $\nu$ itself is an eigenform of $s$ with eigenvalue
$m$. Hence, for any eigenvalue $m^2$ of the Laplacian, $\pm m$ are
eigenvalues of the signature operator. 

More precisely, eigen--forms of $s$ are actually formal sums of forms rather
than forms of a given degree alone; this is what bispinors are. (Also, since
$s$ changes odd forms into even ones and viceversa, any massive eigen--form
has at least two summands.) If we denote such an 
``eigen--formal--sum--of--forms'' as $\sum_k \nu_k$, explicitly we have
\begin{equation}
  \label{eq:sign}
d \nu_k + d^\dagger \nu_{k+2} = m \nu_{k+1} \ .
\end{equation}
We can also use the fact that $*$ commutes with $s$. 
Under genericity assumptions each eigenvalue will be isolated, and hence we
have $\nu_k =* \nu_{6-k}$. (\ref{eq:glmw}, \ref{eq:dftv}) are more specific because they give us the action directly of $d$ on forms. But indeed, let us
consider a case in which there are only 
$\nu_2, \nu_3, \nu_4$. Then we have $d\nu_2 + d^\dagger \nu_4 = m \nu_3$, 
$d\nu_3=m\nu_4$,$d\nu_4=0$. 
The first of these equations can be rewritten as $(1+*) d\nu_2=m\nu_3$. 
The resulting expression
is basically the same as (\ref{eq:glmw}), by giving appropriate names;
in particular $E_0\equiv m$. It is also possible to assemble many of
these eigenforms of the signature operator to get an equation like 
(\ref{eq:dftv}). 

So, it would seem that (\ref{eq:glmw}, \ref{eq:dftv}) are nothing but 
particular cases of restriction to the low--lying spectrum of the signature 
operator. 
However, that does not take into account supersymmetry; it is here that
those forms become less trivial. If on the six--manifold $M$ there is 
a spinor without zeros 
(that is, as reminded in section 3, an SU(3) structure), 
it has been argued in \cite{glmw} that the effective four--dimensional
theory will have $\nn=2$ supersymmetry. (To be sure: the effective theory
 will be most times 
spontaneously broken, so that there will be no supersymmetry--preserving
vacua.)

Giving an SU(3) structure is stronger than giving a metric: it implies one. 
Indeed, in this language a metric 
is an O(6) structure. Since SU(3)$\subset$O(6), an SU(3) structure
determines a metric. 

Now, it seems that the general compactification should use a basis of forms
as follows. One can consider the spectrum of the signature operator. 
Every time we truncate the spectrum keeping only 
eigenvalues $m$ such that $|m| < m_0$, we should check whether, in the linear
span (with constant coefficients) of the forms so obtained, there is 
a SU(3) structure, as in point {\it iv)} above. If there is, we should have
produced a consistent $\nn=2$ theory in four dimensions. Then the conjectures
in \cite{glmw} reviewed and examined in last subsection \ref{sec:basis}
would simply mean
that, in the distorted mirror quintic, one needs to include at least one 
massive eigenvalue in order to get a supersymmetric theory in four 
dimensions. In last subsection  
we have also tried to sketch how one would check explicitly
that at least one massive form is needed.

Even if the argument above for including one massive eigenvalue were right, 
clearly it is a bit unsatisfactory that we have to resort to conjectures
or to painfully explicit methods. So we will examine
two more possible ideas, neither completely satisfying. 

The first is to actually go the other way around: knowing the existence
of an SU(3) structure with certain properties, maybe we can infer 
the existence of forms satisfying some version of (\ref{eq:glmw}). 
For example, if $J$ were itself a massive eigen--form of the Laplacian,
we could simply declare it to be, say, $\omega_1$, taking all the other
$\omega$s to be harmonic two--forms. 

If one knows the intrinsic torsion $W_i$, $i=1\ldots 5$, it is easy to see
when this is the case. In fact, we will use a slightly different formulation
for intrinsic torsion, which is better adapted to this question. 
Namely, we can characterize an SU(3) structure by the covariant derivative
of the spinor, rather than by $dJ$ and $d\Omega$. It can be shown indeed
$\nabla_m \epsilon_+$ (the subscript $_+$ denotes chirality) can always
be put in the form \cite{fmt, gmpt}: 
\begin{equation}
  \label{eq:nabla}
  \nabla_m \epsilon_+= i\tilde q_m \epsilon_+ + i q_{mn} \gamma^n \epsilon_-\ .
\end{equation}
where, as is customary, $\epsilon_-=\epsilon_+^*$. The quantities $\tilde q_m$
and $q_{mn}$ are related to the more usual 
$W_i$s by a change of variables \cite{fmt,gmpt}
\begin{equation} 
  \label{eq:qmnhol} 
  \begin{array}{c}\vspace{.3cm}
  q_{ij} = -\frac{i}{8} W^3_{ij} - \frac18\Omega_{ijk}\bar W_4^k \, ,
  \\ 
  q_{i \bar j} = -\frac i4 \bar{W}^2_{i\bar j} +\frac14 \bar W_1 
g_{i\bar j} \, ,
  \end{array}
\qquad  q_i = \frac i2 (W_5 -W_4)_i
\ .
\end{equation} 
Here $i,j\ldots$ and $\bar i,\bar j,\ldots$ are respectively holomorphic 
and antiholomorphic indices (as opposed to real indices $m,n\ldots$ that
we have used so far).

The fastest way to proceed is now to use the bispinors 
$\epsilon_+ \otimes \epsilon_+^\dagger=
\frac18 e^{-i J}\!\!\!\!\!\!\!
\begin{picture}(10,10)
\put(0,0){\line(1,1){10}}
\end{picture}$ and $\epsilon_+\otimes\epsilon_-^\dagger=-
\frac i8 {\slash\!\!\!\Omega}$.
Applying $\nabla_m$ again to (\ref{eq:nabla}), we can see what are the
conditions for either $J$ or $\Omega$ to be massive eigen--forms of
the Laplacian.

We skip the details of the computation, which is straightforward, and give
the results. In the case of $\Omega$, the conditions are
\[
 i \de_m \tilde q^m - 2 \tilde q_m \tilde q^m - q_{mn} q^{mn} = 
\mathrm{const}\ , \qquad (\de_m -2 \tilde q_m)q^m_{\ n} =0 \ ,
\qquad q_{mi} q^m_{\ j}=0\ .
\]
As above, here $i$ and $j$ are holomorphic indices; 
last equation means a projection
of $q_{mn}q^m_{\ p}$ in the representation $6$ of SU(3) is vanishing.

The conditions for $J$ to be eigenform of the Laplacian read
\[
\de_m  q^m_{\ n}=0\ , \qquad \tilde q_m q^m_{\ n} =0\ , \qquad
q_{mn} q^{mn}=\mathrm{const}\ , \qquad q_{mi} q^m_{\ \bar j} -\frac13 
g^{i\bar j} q_{mi}q^m_{\ \bar j}=0\ .
\]
Again $i$ is holomorphic and $\bar j$ antiholomorphic; the last condition
is really the vanishing of the projection of $q_{mn} q^m_{\ p}$ in the $8$. 
Unfortunately neither of these is satisfied for a half--flat manifold;
these conditions might nonetheless help in other cases. It is easy to see
that the case in which only $W_1$ is present is solution for both. 
(In that case, of course the general computation above is an overkill.) 
A $G_2$ analogue of this idea has appeared in \cite{micu}. Another context
in which similar equations appear is when trying to add Bianchi identity
to conditions coming from supersymmetry. 

The second idea would be the following. As we have seen in last subsection, 
including or not a massive form would lead to SU(3) structures with 
different differential properties. So one could determine easily how 
many massive forms have to be included, if there were methods 
to constrain the possible types of SU(3) structures (complex, symplectic, 
\ldots) just by looking at the topology. 

Unfortunately it seems that such a  systematic method does not exist.
The existence of an SU(3) structure does impose some conditions, but they
are rather mild. Existence of an almost complex structure imposes that the
third Stiefel--Whitney class vanishes \cite{wall}, ${\cal W}_3=0$. 
(This bears no relation to 
the intrinsic torsions $W_i$ above.) Existence of a $(3,0)$--form $\Omega$ 
trivializes the canonical bundle of the almost complex structure, which 
we can write as $c_1=0$. 
But one does not know of a method to determine whether on 
a manifold with a certain topology one can find SU(3) structures with 
certain $W_i=0$. 
For example, it has long been an open problem whether $S^6$,
which notoriously admits almost complex structures, also admits a {\it 
complex} structure. This possibility 
has been excluded for almost complex structures orthogonal with respect
to the round metric \cite{lebrun}, 
then to metrics in its neighborhood \cite{bor}, 
and now finally apparently completely solved by Chern \cite{chern}. 

It would be nice if compactifications of supergravity provided even partial
answers to these questions. The way this could possibly work involves coupling
of massless modes to massive ones. To explain this, we have to take a small
detour. We will claim that the massless sector is in one--to--one 
correspondence with the topology of six--manifolds. But to begin with, 
let us review some standard facts in \cy\ compactifications.
Later we will come back at general SU(3) structure manifolds.

Compactifying IIA string theory on a \cy\ gives an $\nn=2$ action with 
$h^{1,1}$ vectors $t^i$ and $h^{2,1}+1$ hypermultiplets $z^a$. 
The action for the vectors is determined by a prepotential ${\cal F}_0(t)$.
This function can be determined by mirror symmetry methods or by genus zero 
topological amplitudes. There are also other terms in the action which
can be computed by topological amplitudes, this time in higher genus. 
These are terms of the 
form, in superfield notation, ${\cal F}_g(t) (W^2)^g$; ${\cal F}_g(t)$ are
topological amplitudes at genus $g$, and $W$ is the so--called Weyl
superfield, which contains the graviphoton and the (self--dual part of the)
Riemann tensor. (In components, these terms include $R^2 (T^2)^{g-1}$ and
$(R\cdot T)^2 (T^2)^{g-2}$.)

The prepotential (genus 0)
 starts cubic in the $t$'s and then involves exponentials:
${\cal F}_0= D_{ijk} t^i t^j t^k + \sum_{d; m_i} n_{m_i}
\frac1 {d^3} e^{d m_i t^i}$. The symmetric tensor $D_{ijk}$ is 
the triple intersection form
The $m_i$ index the various instantons but this will not be important here. 
At genus 1, we have ${\cal F}_1(t)=(c_2)_i t^i + $ a 
more complicated expression involving exponentials, which counts instantons
of genus 1 (as well as having also a contribution from genus 0; in fact
these higher genus contributions are most elegantly written if we sum them all
together, in terms of Gopakumar--Vafa invariants). From genus 2 
on, ${\cal F}_g$ starts with a constant piece and goes on with instantons.

Now we want to consider a certain infrared limit of the theory. First of all
we want to send to infinity the masses of the massive KK modes. Very roughly,
we can consider that this is controlled by the ``size'' of the 
\cy, $R \equiv (V_6)^{1/6}$. As we have
seen in section \ref{sec:basis}, this will in general be not true: 
some eigenvalues may be significantly smaller than all the others, and 
controlled by some other scale. This was an important point in \cite{glmw}. 
But ``generically'' we can assume it is true, 
and then send $R\to 0$ to decouple the KK modes. This may be puzzling 
because usually it is a large--volume limit
which is thought of as infrared. However, what is really meant by that 
is the volume in string units.  
Indeed, if we want also to get rid of the instanton corrections sketched above,
we see (reinstalling powers of $l_s$ in 
the expressions above) that we need $R/l_s \to \infty$. This 
can also be thought of as the requirement that the curvature 
of the manifold be small with respect to the string scale. As a  
consequence of the previous limits, $l_s\to 0$ even faster 
than $R$, which then also decouples the massive string modes.
Summing up, in the limit $l_s \to 0, l_s/R \to 0$, we remain with a certain
infrared limit of the theory in which string modes decouple and KK masses 
go to infinity. This limit should be familiar from AdS/CFT. 

In this infrared limit
we see that the only information kept by the theory is
\begin{itemize}
\item the massless spectrum; 
\item ${\cal F}_0$ and ${\cal F}_1$.
\end{itemize}
The massless spectrum is determined by $H^2$ and $H^3$. ${\cal F}_0$ is
determined by the intersection matrix $D_{ijk}$; ${\cal F}_1$ by the 
Chern class $c_2$, which is proportional to the Pontrjagin class $p_1$, since
the first Chern class is zero. Going back to the Wall--\u{Z}ubr theorem 
that we have reviewed at the beginning of this section, we see that we are
dealing exactly with the same amount of data. 
So, the infrared limit above of string theory on a \cy\ is in 
one--to--one correspondence with its topology. 

Let us now come back at the less solid SU(3) compactifications. 
We have argued above that the theory we started with, before the limit, 
depends crucially on 
differential geometric properties of the SU(3) structure, as argued
in subsection \ref{sec:general}. After the limit, it only depends on topology. 
We want to suggest that this is in fact a feature of compactifications
on all manifolds of SU(3) structure, not only on \cy s. One reason 
is simply that the previous correspondence seems to be suggestive of
this. We can sketch another reason as follows. If we have obtained our 
SU(3) structure manifold by some surgery--like operation from a \cy\ 
(as for the distorted mirror quintic), then we can obtain the 
four--dimensional effective theory by a gauging of the unbroken $\nn=2$ theory
corresponding to the \cy. (For ${\cal F}_0$, this is not completely 
obvious but it
has been assumed in all the papers so far on the subject, and we will comment
more about it below. For ${\cal F}_1$, one would expect it to work because
it is a quantum effect for a supersymmetric action, but some terms require
numerical conspiracies to work \cite{afmn}, so this should be checked.) 
Gauging gives masses to some fields, that will decouple
in the infrared limit above. For all the other fields, ${\cal F}_0$ and 
${\cal F}_1$ are simply the same as before. So they are still connected
to the topology of the manifold. (This limit theory will in general not be
supersymmetric: the SU(3) structure
which existed in the span of the harmonic and massive forms will go away.
In other words, we not always get a \cy. Usually a limit of a supersymmetric 
theory is still supersymmetric; presumably here the limit may not commute
with taking KK reduction.)

After this long digression, we can come back at our original problem, which was
to give criteria for deciding, given a topology which admits SU(3) structures, 
what will be their differential properties (for example if there is any
integrable complex structure). The idea would be 
to reduce this problem to a physical 
problem. Thanks to the one--to--one correspondence above, we can univoquely
give a massless theory coming from compactification on a given manifold.
In other words, we know that two manifolds will not give the same infrared
limit. Now, the full 
supersymmetric theory (with massive states included) knows about the SU(3)
structure, and also about its integrability properties (for example IIB 
vacua require that the almost complex structure be integrable 
\cite{frey,dall'agata,gmpt}). So we could reformulate the problem as
coupling the massless sector to the massive one in a supersymmetric way.

There are other possible uses to the correspondence above. 
Another of its consequences is that
there is no topological information in the massive states which is not
contained in the massless sector already. This is, in a different guise, 
what we also concluded from Miller's theorem on formality. Maybe one can use
the present methods to give a physical proof of that theorem.
Another feature is that for any collection of data $H^2$, $H^3$, ${\cal F}_0$
and ${\cal F}_1$ one can produce a corresponding six--manifold. This may be
important to construct explicit examples of supersymmetric compactifications. 
Finally, consider again the problem which emerged in previous subsection: 
the fact that we can tell the existence of an almost complex structure 
on a six--manifold, but that we cannot predict whether for example it will 
be complex or not. 

To add some more plausibility to the conjectures of this final part 
we would like to stress a general point: many of
the properties of \cy\ compactifications actually survive in the 
SU(3) structure case, with differential geometry replacing
algebraic geometry. Let us make an example.
If the action has to have $\nn=2$ supersymmetry,
it should have the structure of a $\nn=2$ gauged supergravity. 
The gauging of the hypermultiplet moduli space is 
controlled by a hypermomentum map \cite{bertolini}. It turns out that this
hypermomentum map has a {\it
universal} expression. Its form has been argued with increasing detail
in \cite{vafa, mcgreevy, gmpt}; it is currently being
investigated in \cite{lgw}. A typical component of the momentum map
looks like $dJ$ (on the space of $\Omega$s) or $d\Omega$ (on the space of
$J$'s). It is interesting to note that these momentum
maps have already appeared in considerations by Hitchin \cite{hitchinstable},
which also formalizes what is meant by the space of $J$'s and 
of $\Omega$'s. 
This point is implicitly related to the assumption we made above, that
the theory after introduction of intrinsic torsion has the
same prepotentials -- all the change being in the momentum maps. It is 
most commonly assumed for introduction of fluxes rather than for intrinsic
torsions, but the two are connected by T--dualities.  

{\bf Acknowledgments.}
I would like to thank, for useful comments, discussions or correspondence, 
Paul Aspinwall, Bogdan Florea, Shamit Kachru, Amir Kashani--Poor, 
Liam McAllister, John McGreevy, Xiao Liu, 
Andrei Micu, Ruben Minasian, Daniel Waldram, Alberto Zaffaroni,
and especially Mark Gross. This research has been supported in part by the DOE 
under contract DEAC03-76SF00515 and by the NSF under contract 9870115.

\end{document}